\documentclass[prd,preprint,nofootinbib,superscriptaddress]{revtex4-1}

\usepackage{graphicx}
\usepackage{amssymb}
\usepackage{amsmath}
\usepackage{color}

\IfFileExists{srcltx.sty}{\usepackage[active]{srcltx}}

\begin{document}
\renewcommand{\thefootnote}{\#\arabic{footnote}}
\newcommand{\rem}[1]{{\bf [#1]}}
\newcommand{\gsim}{ \mathop{}_ {\textstyle \sim}^{\textstyle >} }
\newcommand{\lsim}{ \mathop{}_ {\textstyle \sim}^{\textstyle <} }
\newcommand{\vev}[1]{ \left\langle {#1}  \right\rangle }
\newcommand{\bear}{\begin{array}}  
\newcommand {\eear}{\end{array}}
\newcommand{\bea}{\begin{eqnarray}}   
\newcommand{\eea}{\end{eqnarray}}
\newcommand{\beq}{\begin{equation}}   
\newcommand{\eeq}{\end{equation}}
\newcommand{\bef}{\begin{figure}}  
\newcommand {\eef}{\end{figure}}
\newcommand{\bec}{\begin{center}} 
\newcommand {\eec}{\end{center}}
\newcommand{\non}{\nonumber}  
\newcommand {\eqn}[1]{\beq {#1}\eeq}
\newcommand{\la}{\left\langle}  
\newcommand{\ra}{\right\rangle}
\newcommand{\ds}{\displaystyle}
\newcommand{\red}{\textcolor{red}}
\def\SEC#1{Sec.~\ref{#1}}
\def\FIG#1{Fig.~\ref{#1}}
\def\EQ#1{Eq.~(\ref{#1})}
\def\EQS#1{Eqs.~(\ref{#1})}
\def\lrf#1#2{ \left(\frac{#1}{#2}\right)}
\def\lrfp#1#2#3{ \left(\frac{#1}{#2} \right)^{#3}}
\def\GEV#1{10^{#1}{\rm\,GeV}}
\def\MEV#1{10^{#1}{\rm\,MeV}}
\def\KEV#1{10^{#1}{\rm\,keV}}
\def\REF#1{(\ref{#1})}
\def\lrf#1#2{ \left(\frac{#1}{#2}\right)}
\def\lrfp#1#2#3{ \left(\frac{#1}{#2} \right)^{#3}}
\def\OG#1{ {\cal O}(#1){\rm\,GeV}}

\begin{flushright}
CRR-Report-607-2011-24 \\
IPMU 12-0021 
\end{flushright}

\title{
Primordial seeds of supermassive black holes}
\author{Masahiro Kawasaki}
\affiliation{Institute for Cosmic Ray Research, The University of Tokyo,
5-1-5 Kashiwanoha, Kashiwa, Chiba 277-8582, Japan}
\affiliation{Institute for the Physics and Mathematics of the Universe,
University of Tokyo, Chiba 277-8583, Japan}
\author{Alexander Kusenko}
\affiliation{Institute for the Physics and Mathematics of the Universe,
University of Tokyo, Chiba 277-8583, Japan}
\affiliation{Department of Physics and Astronomy, University of California, Los
Angeles, CA 90095-1547, USA}
\author{Tsutomu T. Yanagida}
\affiliation{Institute for the Physics and Mathematics of the Universe,
University of Tokyo, Chiba 277-8583, Japan}

\begin{abstract}

Supermassive black holes exist in the centers of galaxies, including Milky Way, but there is no compelling theory of their formation. Furthermore, observations of quasars imply that supermassive black holes have already existed at some very high redshifts, suggesting the possibility of their primordial origin.  In a class of well-motivated models, inflationary epoch could include two or more periods of inflation dominated by different scalar fields. The transition between such periods of inflation could enhance the spectrum of density perturbations on some specific scale, which could lead to formation of primordial black holes with a very narrow range of masses of the order of $10^5$ solar masses.
These primordial black holes could have provided the requisite seeds for the observed population of supermassive black holes.

\end{abstract}


\maketitle

\newpage

\section{Introduction}

Supermassive black holes (SMBHs) with masses $10^6-10^{9.5} M_\odot$ reside in the centers of most galaxies~\cite{Kormendy:1995er,astro-ph/9708072,astro-ph/9810378}, 
including Milky Way~\cite{Eckart:1996zz,astro-ph/9807210,astro-ph/0306130}.  Furthermore, observations of quasars (QSO) with redshifts $z>6$ imply that SMBHs must have already existed at such high redshifts.  
Stellar explosions can produce black holes with masses up to $10-15 M_\odot$~\cite{astro-ph/9911312}, but there is no mechanism by means of which such small objects could grow to become 
SMBHs~\cite{astro-ph/0404196}, except if dark matter has a sufficient self-interaction to facilitate a rapid transfer of angular momentum and kinetic energy~\cite{astro-ph/0108203} (such models exist, but they are somewhat {\em ad hoc}~\cite{astro-ph/9909386,astro-ph/0106008}). 
The early formation of SMBHs, which is necessary to account for high-redshift quasars, implies that SMBHs may have preceded star formation~\cite{astro-ph/9801013}.
The masses of SMBHs exhibit a remarkable correlation with the bulge masses of their host galaxies~\cite{astro-ph/9708072,astro-ph/9801013,astro-ph/0107134,Graham:2012bs}.  The bulge mass is $10^3$ times larger than the black hole mass, and the proportionality holds over some four orders of magnitude.   

\begin{figure}
\begin{center}
\includegraphics[width=0.4\textwidth]{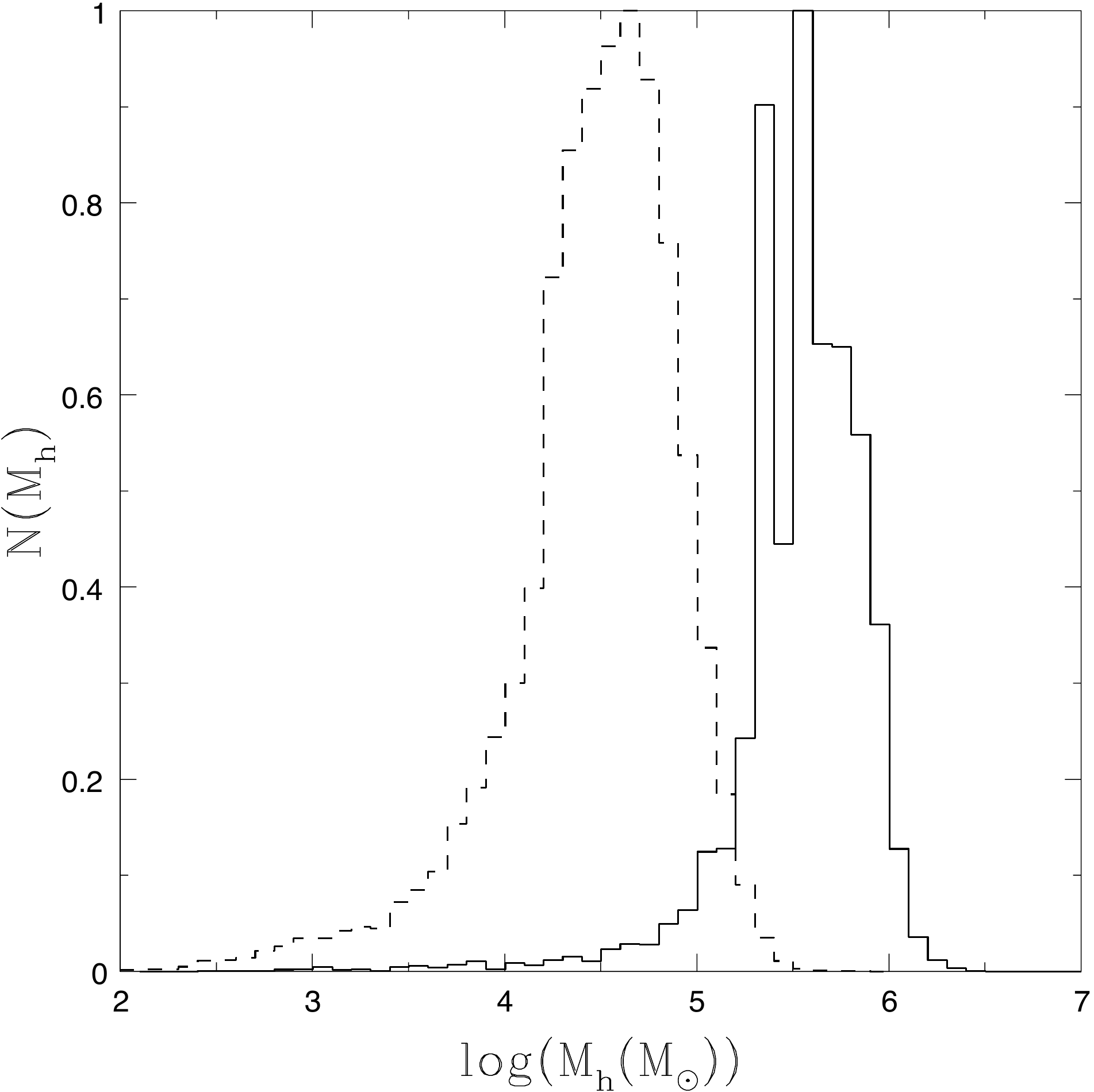}\includegraphics[width=0.4\textwidth]{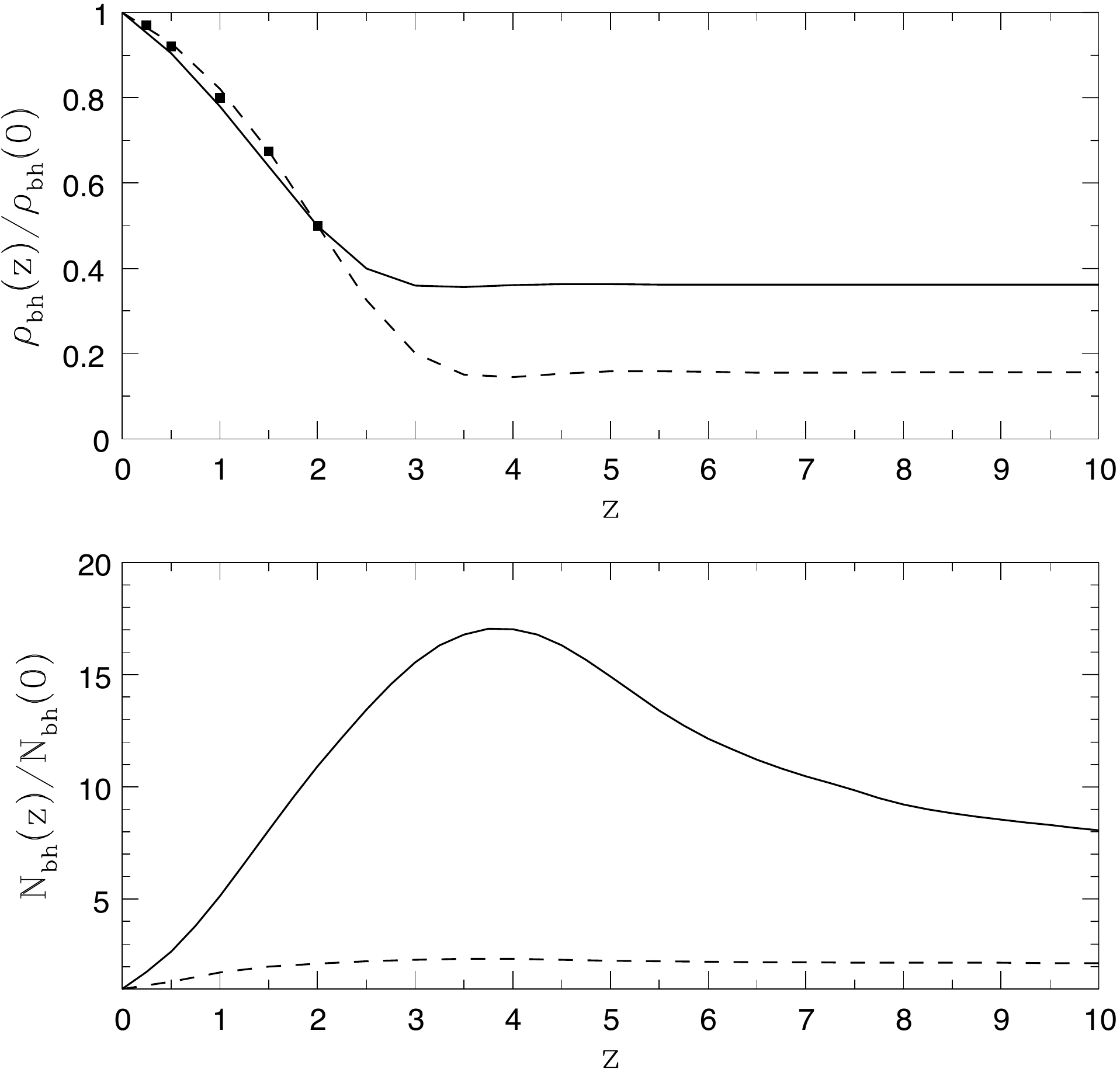}
\caption{The devolved mass distribution of PBHs (on the left) that produces an acceptable evolution of mass and number density of black holes in 
galactic halos (right), according to Bean and Magueijo~\cite{astro-ph/0204486}.  The data points are from Ref.~\cite{1992MNRAS.259..421C}.  The two distributions are for the limiting halo mass of 
$10^9M_\odot$ (dashed line) and $10^{10} M_\odot$ (solid line); see discussion in Ref.~\cite{astro-ph/0204486}. 
\label{fig:Bean}}
\end{center}
\end{figure}

The robust proportionality between the mass of the SMBH and the mass of the bulge, and the early QSO activity suggest that the origin of SMBH may be primordial.  
A number of processes in the early universe could lead to formation of primordial black holes~\cite{Khlopov:2008qy}. Collapse of primordial density fluctuations~\cite{Carr:1975qj} usually results in a broad distribution of masses, unless some specific scale is favored, for example, by a first-order phase transition~\cite{Jedamzik:1998hc}, or by formation of domain walls in a second-order phase transition~\cite{Rubin:2001yw}.   The idea of primordial origin of SMBHs was examined in Ref.~\cite{astro-ph/0406260} for the class of models that generate a broad, power-law spectrum of black hole masses.  In such models, the required number of SMBH seeds can only be produced at the expense of generating too many smaller-mass black holes, which leads to a conflict with existing upper bounds.  
Furthermore, aside from observational bounds, the history of black hole formation and mergers favors a rather narrow mass function for primordial seeds.  Bean and Magueijo~\cite{astro-ph/0204486} have devolved the black hole mass function inferred from observations and obtained the required distribution of seed masses, shown in Fig.~\ref{fig:Bean}.  This distribution, is extremely narrow, much more so than the distributions expected from most models of the early universe.  

However, it was recently pointed out that, in a class of hybrid double inflation models, a very narrow distribution of primordial black hole (PBH) masses can be 
produced~\cite{hep-ph/9710259,Yokoyama:1998pt,Kawasaki:2006zv,Kawasaki:2007zz,Kawaguchi:2007fz,Saito:2008em,arXiv:1001.2308}.  Furthermore, the masses of PBH produced by inflation can 
lie in a very broad range, and specific models have considered masses from as small as $10^{-7} M_\odot$ and as large as $10^{9} M_\odot$, including a particularly interesting range around $0.1 M_\odot$, which could be probed by gravitational lensing of massive compact objects in the Milky Way halo~\cite{Yokoyama:1995ex,Kawasaki:2006zv}.    
Scalar fields with flat potentials are generic predictions of many theories beyond the standard model, 
and a two-stage or multi-stage inflation can actually be a generic feature of supersymmetry and string theory~\cite{Dvali:2003vv,Burgess:2005sb}. 
String theory predicts a large number of scalar fields, moduli, such as compactification radii, which appear as low-mass fields in the low-energy effective field theories.  Supersymmetry 
(in theories with or without superstring UV completion) predict a large number of flat directions in the potential.  Some flat directions are parameterized by scalar fields that gain 
relatively small masses from supersymmetry breaking.  The light scalars present a very difficult problem for cosmology~\cite{Coughlan:1983ci,hep-ph/9308325} because their coherent oscillations can come to 
dominate the energy density of the universe very late.  With the help of thermal inflation~\cite{Lyth:1995hj,Lyth:1995ka}, or by means of coupling the moduli fields to the inflaton~\cite{Linde:1996cx,Nakayama:2011wqa}, the moduli problem can be ameliorated in some range of masses~\cite{Hashiba:1997rp,Asaka:1997rv,Asaka:1998ju}, but neither approach provides a complete solution.  However, some scalar fields with relatively flat potentials can serve as inflatons and allow for a cascade of sequential stages of inflation. 
Each of these stages may be too short to account for the observed flatness and homogeneity of the universe, but they can be responsible for generating primordial black holes that can play the role of dark matter~\cite{Asaka:1998ju,arXiv:1001.2308}.  The moduli fields can generate density perturbations~\cite{Kofman:2003nx}. Given the plethora of light scalar fields expected in the landscape of string vacua, a 
multiple-stage inflationary cascade appears to be a rather generic possibility.  We will concentrate on the latest two stages of this potentially elaborate cascade. 

Since it is likely that more than one inflationary epoch could have taken place in the early universe, the slope of the scalar potential could have an abrupt change 
in transition from one inflation to another.  
The transition between inflations can enhance the spectrum of density perturbations on some specific scale.  When these density perturbations re-enter the horizon, they can form a population of black holes with masses in a narrow range.  The masses produced in these models can be as large as $10^5 M_\odot$, which is consistent with the range required for seeding SMBHs shown in Fig.~\ref{fig:Bean}.  This range is also consistent with the observational constraints.  Assuming that every galaxy has a seed PBH of mass $\sim (10^4 - 10^5) M_\odot$, the fraction of dark matter in such objects is below $10^{-5}$, which is below the observational upper limits~\cite{Ricotti:2003vd,Ricotti:2007au}.

\section{PBH formation in a double inflation model}

Here we describe the basic idea how a double inflation model produces the density perturbation spectrum with a sharp peak. 
After inflation an inflation field begins to oscillate.  During this oscillating phase, if the inflation $\phi$ has an appropriate 
coupling with another field $\psi$, it decays rapidly into $\psi$ through parametric resonance.
Because of the band structure of the parametric resonance, the inflation dominantly decays into $\psi$ particles with a specific momentum $k_p$.
In terms of fields, the parametric resonance induces large fluctuations of the $\psi$ field in a very narrow wavelength range. 
In general, the enhanced $\psi$ fluctuations have a momentum of the order of the inflation mass $m_\phi$, which corresponds to length scales smaller than the horizon size, 
and, therefore, they do not affect the density perturbations on cosmological scales.

However, if there is a second stage of inflation following the first one, the situation changes. 
Let us suppose that the second inflation is caused by some scalar field $\varphi$.
When the second inflation $\varphi$ has a coupling with $\psi$ field, the enhanced $\psi$ fluctuations become a source of $\varphi$ fluctuations during oscillating phase of the first inflation. 
As a result, $\varphi$ has large fluctuations at $k\simeq k_p$.  
Since these fluctuations are exponentially extended by the second inflation, they generate cosmologically relevant density perturbations whose spectrum has a sharp peak. 
This mechanism, which produces a sharp peak in the spectrum of density perturbations, was first considered in the supersymmetric smooth hybrid new inflation model~\cite{Kawasaki:2006zv} (see also~\cite{Kawasaki:2007zz,Kawaguchi:2007fz,arXiv:1001.2308})  and then in K\"ahher moduli double inflation~\cite{Kawasaki:2010ux}. 
Evidently, the mechanism operates in a variety double inflation models.

Now let us consider the relation between BH mass and the parameters of double inflation. The PBHs are produced when the density perturbation re-enters the 
horizon, $k/a = H$, and their average mass is approximately given by the horizon mass. 
Since the spectrum has a sharp peak at $k=k_p$, produced PBHs also have a specific mass $M_{\rm BH}$ corresponding to $k_p$. $M_{\rm BH}$ and $k_p$ are related: 
\begin{eqnarray}
    M_{\rm BH} & \simeq & 2\times 10^{13}~M_{\odot} 
    \left(\frac{k_p}{{\rm Mpc}^{-1}}\right)^{-2}, 
    \label{eq:PBH_mass}  \\[0.5em]
    k_p & \simeq & 4.6\times 10^6 ~{\rm Mpc}^{-1} 
        \left(\frac{M_{\rm BH}}{M_{\odot}}\right)^{-1/2}. 
\end{eqnarray}
Here $k_p$ is a comoving wavenumber.
The peak wavenumber $k_p$ is of the order of the inflation mass,
\begin{equation}
  \frac{k_p}{a_{\rm osc}} ~= ~\alpha m_{\phi},
\end{equation}
where $a_{\rm osc}$ is the scale factor at the start of the $\phi$ oscillation after the first inflation and $\alpha = O(0.1)$. 
Let us define $t_*$ and $a_{*}$ as the time and the scale factor when the pivot scale ($k_{*} = 0.0002$~Mpc$^{-1}$) leaves the horizon during the first inflation. 
$a_{*}$ is determined by  
\begin{equation}
   \frac{k_{*}}{a_{*}} = H_{1},
\end{equation}
where $H_1$ is the Hubble parameter during the first inflation. 
Then, the number of e-folds $N_{1*}$ from $t_{*}$ to the end of the first inflation is
\begin{equation}
    N_{1*} \simeq 21.6 + \ln \left(\frac{H_1}{\alpha m_{\phi}}\right) 
    - \frac{1}{2}\ln\left(\frac{M_{\rm BH}}{M_{\odot}}\right).
    \label{eq:efold_2nd_inf}
\end{equation}
$N_{1*} \sim 14$ for $M_{\rm BH}= 10^{6}M_{\odot}$ and $\ln (H_1/\alpha m_\phi)\sim 1$.
The pivot scale corresponds to $N_{*} = 50-60$, so the difference  $N_{*} - N_{1*}$ should be provided by the second inflation. 

Besides the sharp peak, the double inflation predicts some characteristic features in the spectrum of density perturbations. 
Because perturbations on large scales and those on small scales are produced during the first and second inflations respectively, their amplitudes are generally different. 
The amplitude on large scales is fixed by CMB observations~\cite{Komatsu:2010fb}, but the density perturbations on small scales can be much larger or smaller depending on models of the second inflation. 
In addition, the double inflation predicts a relatively large running of the spectral index $d n_s/\ln k$ where $n_s$ is the spectral index.  
In single-field slow-roll inflation, $d n_s/\ln k$ is negligible at observable scales because of smallness of slow-roll parameters.  
However, in the double inflation the observable scales exit the horizon near the end of the first inflation when the inflaton rolls down the potential faster, which leads to increase of running of the spectral index.

\section{Smooth Hybrid New Inflation Model}

In this section, we discuss an example of a double inflation models which produces a 
sharp peak in the spectrum of density perturbations. 
The model we adopt here is the smooth hybrid new inflation model~\cite{Kawasaki:2006zv}. 
The original model was built in the framework of supergravity, but the essence of the inflationary dynamics 
is described by the following scalar potential reduced from the full scalar potential in supergravity: 
\begin{eqnarray}
  V    &= & V_{\rm H} + V_{\rm N} + V_{\rm HN}, \\[0.5em]
  V_{\rm H}(\phi, \psi) & = & \left(1+\frac{\phi^4}{8} + \frac{\psi^2}{2}\right)
  \left( -\mu^2 + \frac{\psi^4}{16M^2}\right)^2 +\frac{\phi^2\psi^6}{16M^4}, 
  \\[0.5em]
  V_{\rm N}(\varphi) & = & v^4 \left(1-\frac{\kappa}{2}\varphi^2\right)
  -\frac{g}{2}v^2\varphi^4 + \frac{g^2}{16}\varphi^8, \\[0.5em]
  V_{\rm HN}(\phi,\psi,\varphi) & = & 
  \left( -\mu^2 + \frac{\psi^4}{16M^2}\right)^2\frac{\varphi^2}{2}
  -\left(-\mu^2 + \frac{\psi^4}{16M^2}\right)v^2\phi\varphi, 
\end{eqnarray}
where $\phi$ and $\psi$ are the inflation and waterfall fields of the first inflation ($=$ smooth hybrid inflation~\cite{Lazarides:1995vr}), $\varphi$ is the inflation field of the second inflation ($=$ new inflation~\cite{Izawa:1996dv}), $\mu$ and $v$ are the scales of the smooth hybrid and the new inflations ($\mu > v$), $M$ is some cut-off scale, and $\kappa$ and $g$ are coupling constants. 
Here we set $M_p =2.4\times 10^{18}~{\rm GeV} = 1$.
The scalar potential consists of three parts, $V_{\rm H}$, $V_{\rm N}$ and $V_{\rm HN}$ which are responsible for smooth hybrid inflation, new inflation and their interaction, respectively. 

For $\phi \gtrsim \sqrt{\mu M}$, the potential $V_{\rm H}$ has a local minimum 
at $\psi \simeq \frac{2\mu M}{\sqrt{3}\phi}$, where the potential is described  as
\begin{equation}
  V_{\rm H} \simeq \mu^4 \left(1 +\frac{\phi^4}{8} 
  - \frac{2\mu^2 M^2}{27\phi^4} + \dots\right).
\end{equation}
This potential is dominated by the vacuum energy $\mu^4$ and smooth hybrid inflation 
takes place. 
The Hubble parameter of the smooth hybrid inflation is given by $H_1 = \mu^2/\sqrt{3}$.
During the period of smooth hybrid inflation, the potential $V_{\rm HN}$ can be written as
\begin{equation}
   V_{\rm HN} = \frac{1}{2}\mu^4\varphi^2 + \mu^2v^2\phi\varphi +\ldots .
\end{equation}
Thus, the inflaton of the new inflation is stabilized at $\varphi \simeq - (v/\mu)^2\phi$, which evades the initial value problem for new inflation.  

After inflation, the inflaton field $\phi$ and the waterfall field $\psi$ rolls down toward the true minimum $\phi=0$ and $\psi = 2\sqrt{\mu M}$ and start oscillation.
During the oscillating phase, fluctuations of both fields, $\delta\phi$ and $\delta\psi$ grow rapidly through parametric resonance due to self- and mutual couplings.   
The fastest growing mode has wave number 
\begin{equation}
   \frac{k_p}{a_{\rm osc}} \simeq 0.3 m_\phi,
   \label{eq:peak_wave_number}
\end{equation}
where $m_\phi = \sqrt{8\mu^3/M}$ is the mass of the inflation $\phi$.
Then, the fluctuations $\delta\phi$ and $\delta\psi$ amplify the fluctuations of the inflation $\varphi$ of the new inflation via the interaction represented by the second term in $V_{\rm HN}$.  
The most amplified mode has the same wave number as that given by Eq.~(\ref{eq:peak_wave_number}).
After $\phi$ and $\psi$ decay by parametric resonance and usual perturbative process,  the vacuum energy $v^4$ in $V_{\rm N}$ dominates the universe and the second stage of inflation, i.e. new inflation starts.  
As a result, the enhanced fluctuations of $\varphi$ produce a sharp peak in the spectrum of the curvature perturbations.

\begin{figure}[t]
\begin{center}
\includegraphics[width=0.7\textwidth]{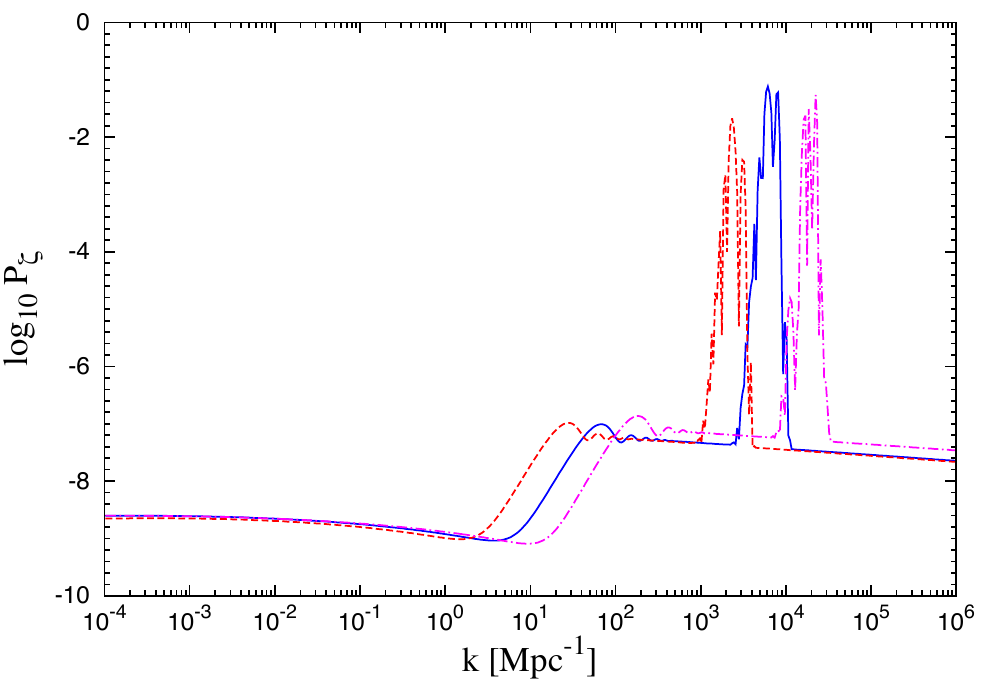} 
\caption{Power spectra of curvature perturbations for $\mu \simeq 2 \times 10^{-3}, v=\mu/4,  \kappa = 0.05$ and $g=2\times 10^{-5}$. The solid, dashed and dashed dotted  curves are for $M= 0.52, 0.59$ and $0.48$, and their peaks are responsible for formation of PBHs with mass $\simeq 5\times 10^{5}M_{\odot}, 5\times 10^{6}M_{\odot}$ and  $5\times 10^{4}M_{\odot}$, respectively.
\label{fig:spectrum}}
\end{center}
\end{figure}

We obtain the power spectrum by numerically solving the evolution equations for linear perturbations of the fields and metric.\footnote{
In the calculation we do not include non-linear effects and back-reactions which may be important when the parametric resonance occurs. 
However, some part of back-reaction is taken into account by taking large decay rate for the inflation and waterfall fields of the smooth hybrid inflation.}
The resultant power spectra of the curvature perturbations, ${\cal P}_{\zeta}(k)$, are shown in Fig.~\ref{fig:spectrum}.  
In obtaining the spectra we take $\mu \simeq  2\times 10^{-3}, v=\mu/4,  \kappa = 0.05$ and $g=2\times 10^{-5}$. 
The solid, dashed and dashed dotted  curves are for $M= 0.52, 0.59$ and $0.48$, and they have narrow peaks at $k \simeq 6\times 10^3$~Mpc$^{-1}$, $2\times 10^3$~Mpc$^{-1}$ and $2\times 10^4$~Mpc$^{-1}$ which correspond to PBH masses $\simeq 5\times 10^{5}M_{\odot}$, $5\times 10^{6}M_{\odot} $ and $5\times 10^{4}M_{\odot}$, respectively [Eq.~(\ref{eq:PBH_mass})].
The substructure of the peaks is due to resonance bands. 
In Fig.~\ref{fig:spectrum} the curvature perturbations on large scales ($k \lesssim 1$--$10$~Mpc$^{-1}$) are produced during the smooth hybrid inflation and their amplitudes and spectral indices are consistent with WMAP 7-year data ${\cal P}_\zeta(k_*) = (2.43\pm 0.11)\times 10^{-9}$ and $n_s = 0.963\pm 0.014$~\cite{Komatsu:2010fb}.
On the other hand, the new inflation produce the curvature perturbations on small scales ($k \gtrsim 1$--$10$~Mpc$^{-1}$) and the amplitude is much larger. 
Since the observed spectrum is almost flat at $k \lesssim 1$~Mpc$^{-1}$, the new inflation should only account for the perturbations at $k \gtrsim 1$~Mpc$^{-1}$. 
This sets the upper limit on the e-folding number of the new inflation and  the mass of PBHs from Eq.~(\ref{eq:efold_2nd_inf}) which is given by $M_{\rm BH} \lesssim  {\rm a~few}\times 10^{6}M_{\odot}$. 

A more stringent constraint on the amplitude of the power spectrum comes from the CMB spectral distortion due to photon diffusion~\cite{Hu:1994bz}. 
After the perturbations re-enter the horizon, fluctuations in photon-baryon fluid starts acoustic oscillation. 
During acoustic oscillation, photon diffusion dissipates fluctuations on small scales. 
Then the energy of the acoustic oscillation of the photon fluctuations is injected into the background radiation and distorts the CMB spectrum unless the injected energy is thermalized by photon number changing processes such as double Compton scattering. 
For a given redshift $z$ the critical scale $k_{\rm d}^{-1}$ below which the fluctuations are damped is 
\begin{equation}
    k_{\rm d}^{-1} \simeq 2.5\times 10^5(1+z)^{-3/2}~ {\rm Mpc}.
\end{equation}
When the double Compton scattering becomes ineffective at $z_{\rm DC} \simeq 2\times 10^6$, the critical scale is given by $k_{\rm d}(z_{\rm DC}) \simeq 10^{4}~{\rm Mpc}^{-1}$.
Thus, the photon fluctuations with  $k \lesssim k_{\rm d}(z_{\rm DC})$ dissipate after $z_{\rm DC}$ and cause the spectral distortion which is parametrized by the chemical potential $\mu$.
From the COBE observation  $\mu$ should be $\lesssim 10^{-4}$~\cite{Fixsen:1996nj}.
In our model $\mu$ is given by $\sim \int^{k_{\rm d}} d\ln k P_{\zeta}(k)$ which amounts to $O(10^{-2})$ if $k_p \lesssim k_{\rm d}$.
So we must require $k_p \gtrsim k_{\rm d}$, which leads to the stringent upper bound on the PBH mass as $M_{\rm BH} 
\lesssim 10^5M_{\odot}$.
Therefore, the present model can produce PBHs with mass $10^{4-5}M_{\odot}$ which evolve to SMBHs.

\begin{figure}
\begin{center}
\includegraphics[width=0.7\textwidth]{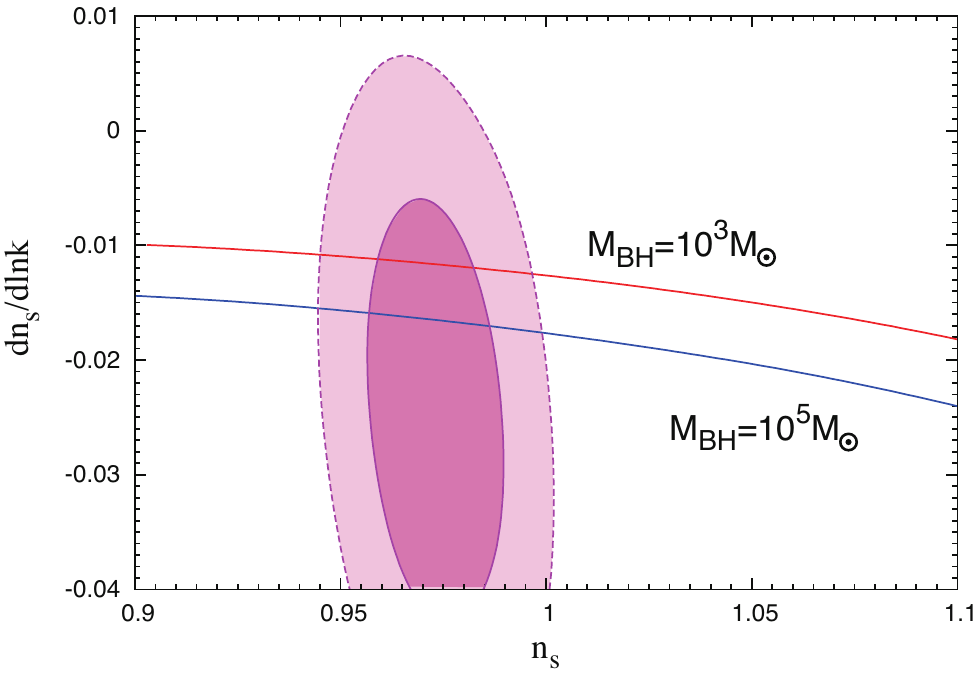} 
\caption{Spectral index and the running parameter predicted in the smooth hybrid new inflation model for $M_{\rm BH} = 10^3M_{\odot}$ and $10^5M_{\odot}$. We also show the constraint from the combined analysis of WMAP and SPT data~\cite{arXiv:1105.3182}
\label{fig:index}}
\end{center}
\end{figure}

As mentioned in the previous section, double inflation models predict a large running spectral index $d n_s/d\ln k$. 
In Fig.~\ref{fig:index} the contours for PBH mass $10^{3}M_{\odot}$ and $10^{5}M_{\odot}$ are shown in the $n_s$--$d n_s/d\ln k$ plane. 
The spectral running is more significant for larger PBH mass because the observable scales exit horizon at later epoch of the first inflation when the slow-roll parameters are larger. 
The predicted spectral running is consistent with the recent CMB observations by SPT~\cite{arXiv:1105.3182} and ACT~\cite{Dunkley:2010ge} which suggest negative spectral running. 
If the spectral index and its running are determined   more precisely, it would be a good test for the PBH formation mechanism by double inflation.

\section{CMB and reionization}

Primordial black holes can absorb gas and produce ionizing X-ray radiation, which can have observable consequences~\cite{Ricotti:2003vd,Ricotti:2007au}. X-rays emitted by supermassive primordial black holes can produce ionization and heating of gas during the ``dark ages''.  While direct ionization is probably too weak to affect the optical depth measured by WMAP, the increase is significant enough to enhance formation of molecular hydrogen, which is an important cooling agent needed for the collapse of gas clouds leading to formation of the first stars.  WMAP constraints~\cite{Ricotti:2007au} allow for one supermassive primordial black hole per galaxy, as our model predicts.  The effects of increased H$_2$ fraction on star formation~\cite{Ricotti:2003vd} can be probed by future 21-cm observations. 

\section{Conclusion}
In this Letter we have considered primordial black holes (PBHs) as seeds for supermassive black holes (SMBHs), 
and we studied their formation in double inflation models.
To seed SMBHs with masses $10^{9-10}M_{\odot}$, PBHs should have masses as large as $10^{4-5}M_{\odot}$. 
We have investigated the smooth hybrid new inflation model as a model of double inflation and numerically solved the evolution of inflaton fields and their fluctuations.  
It was shown that the smooth hybrid new inflation generates a spectrum of the density perturbations with a sharp peak which leads to formation of PBHs with masses $\sim 10^{5}M_{\odot}$. 
The sharp peak in the spectrum ensures that the resultant mass distribution for PBHs is also very narrow, which is consistent with the mass function inferred from observation~\cite{astro-ph/0204486}. 

We have adopted a specific model for double inflation in the present Letter. 
However, double or multiple-stage inflation is quite generic in supersymmetry and string theory, 
and inflaton decays due to parametric resonance in preheating epochs between inflationary stages are also common. 
Therefore, it can be expected that PBHs with a narrow mass distribution are produced in the inflationary universe. 
Besides the sharp peak in the density perturbation spectrum, double inflation models generally predict a large running of the spectral index. 
For the smooth hybrid new inflation model we obtain  $ -0.02 \lesssim (dn_s/d\ln k) \lesssim -0.01$, which can be tested in future CMB observations.    

We thank E.~Pajer and N.~Yoshida for helpful comments and discussions.  
This work was supported in part by DOE grant DE-FG03-91ER40662 (A.K.), 
by Grant-in-Aid for Scientific Research  4102004 and 21111006 (M.K.) 
from the Ministry of Education, Culture, Sports, Science and Technology in Japan, 
and also by World Premier International Research Center Initiative (WPI Initiative), MEXT, Japan.

\end{document}